\documentclass{vgtc}                          

\graphicspath{{figures/}{pictures/}{images/}{./}} 

\usepackage{times}                     

\usepackage{tabu}                      
\usepackage{booktabs}                  
\usepackage{lipsum}                    
\usepackage{mwe}                       
\usepackage{svg}
\usepackage{mathptmx}                  
\usepackage{subcaption}

\usepackage{booktabs}
\usepackage{pifont}  

\onlineid{0}

\vgtccategory{Research}

\vgtcinsertpkg

\title{FluidViews: Adaptive Drag‑and‑Drop Token Filters for Heterogeneous Multi‑View Visual Analytics }

\author{Bhanu Sunku}
\affiliation{Northern Illinois University \\ \href{mailto:z1974769@students.niu.edu}{z1974769@students.niu.edu}}

\teaser{
  \centering
\includegraphics[width=\textwidth, height=4cm]{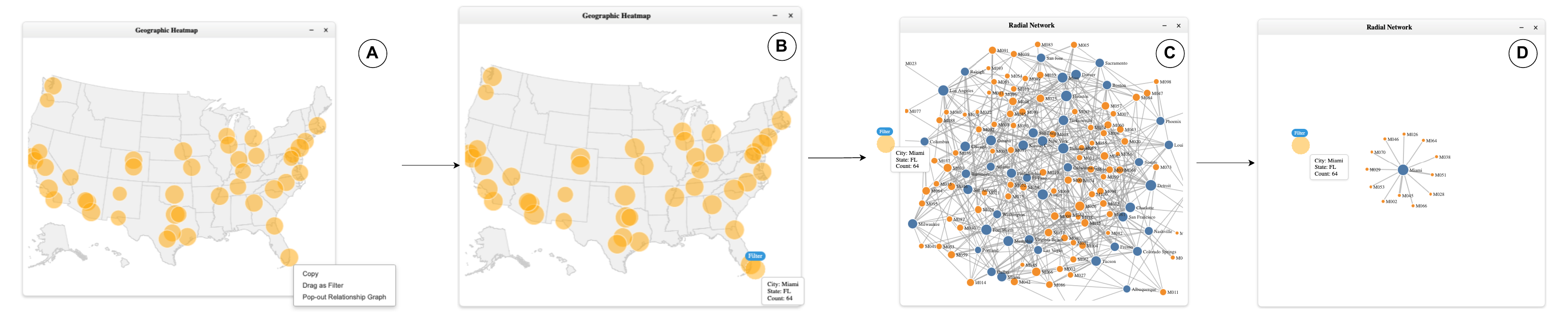}  \caption{FluidViews’ drag-as-filter interactions. (A) The user initiates a direct-manipulation gesture by right-clicking a geographic data point and selecting “Drag as Filter.” (B) A clone of the selected mark is dragged across the interface. (C) Dropping the token onto the radial network view triggers a context-aware filter, highlighting related entities. (D) A pop-out micro-view reveals first-hop connections for the filtered element, supporting detailed exploration without disrupting the main workspace.}
  \label{fig:teaser}
}

\abstract{
   Interactive visual analytics workflows are often disrupted by rigid filter panels and context switches that break analysts’ cognitive flow. We introduce \textbf{FluidViews}, a web-based framework that elevates filters to first-class, manipulable objects through two novel direct-manipulation interactions. \emph{Copy-as-Highlight} enables users to duplicate any visual mark into a persistent highlight token for rapid, transient cross-view comparison, while \emph{Drag-as-Filter} allows analysts to pick up a mark and drop it onto another view to apply context-sensitive filters in place no menus, panels, or modal dialogs required. An optional pop-out micro-view provides on-demand, spatially independent subviews for detailed inspection without disrupting the primary workspace. By embedding these lightweight gestures into coordinated multi-view environments, FluidViews preserves analytic momentum, reduces cognitive overhead, and supports fluid, multi-step exploration across heterogeneous datasets. We describe the system’s design and implementation, illustrate its application in exploratory workflows, and discuss how tangible filter objects can transform interactive data exploration.
} 


\begin{document}


\firstsection{Introduction}

\maketitle


\textit{Coordinated multiple-view visualizations} (MVs) are a central strategy for supporting exploratory data analysis across complex, multi-dimensional datasets. \cite{north2000snap, Baldonado2000Guidelines, roberts2007state}.
By juxtaposing different representations, MV systems help users detect relationships, form hypotheses, and reason across diverse facets of data \cite{north2000snap}. 
To support such cross-view analysis, users must often apply dynamic filters, isolating relevant subsets, focusing on anomalies, or comparing conditions across views.
However, the design of filtering interactions in conventional MV systems remains limited.

Conventional approaches often delegate filtering to static side panels, faceted search widgets, or modal dialogs that are separate from the views themselves, fragmenting user attention and analytic flow \cite{stolte2008polaris, hearst2006design}.
Moreover, many systems lack mechanisms for constructing filters directly through interactions within the views. 
Instead, they often limit cross-view coordination to brushing and linking, where transient, hover-based highlights provide only momentary associations without supporting persistent filtering or multi-condition query building \cite{Baldonado2000Guidelines, song2010comparative}.
As a result, users are forced to shift focus between views, leading to disjointed analysis experiences and an increased cognitive burden when composing or recalling complex filter states \cite{Sifer2006FilterCF, Hernandez2008SynchronizedTC}.

Despite the importance of fluid, tightly integrated filtering in coordinated multiple-view systems, existing techniques offer limited support for constructing, combining, and managing filters directly within the visualization space \cite{Baldonado2000Guidelines, heer2007animated}.
This gap raises critical questions for multi-view interaction design: 
How might we make filters tangible and manipulable within and across views to support more direct, expressive analysis workflows?
How might we enable users to fluidly compose, persist, and transfer filters across coordinated views without disrupting analytic flow?
Addressing these challenges requires rethinking filters not as static menu configurations, but as first-class, interactive objects embedded within the analytic workspace. 

To address these challenges, we present \textit{FluidViews}, a web-based visual analytics framework. that enables seamless, expressive interaction across coordinated views. Our contributions are twofold: (1) we introduce a novel interaction framework, \textit{Direct Filter Manipulation} (DFM), that treats visual marks as manipulable, first-class objects embedded directly within the analytic workspace, allowing users to fluidly construct, transfer, and persist analytic context without relying on detached control panels; (2) we implement this framework in a proof-of-concept system that supports a suite of complementary techniques, including \emph{drag-as-filter}, which allows users to drop visual marks onto other views to apply filters; \emph{copy-as-highlight}, which propagates selections across views for transient comparison; hover-based multi-facet highlighting with click-to-lock persistence; and pop-out micro-views that enable detailed inspection without disrupting the main analytic flow. Together, these techniques extend DFM principles to support a more fluid, multi-step exploration experience across diverse visualizations.

\section{Background Work}

\subsection{Coordinated Multiple Views }
Coordinated Multiple views have long been used to provide different perspectives on complex data, allowing users to explore relationships via interactive linking and brushing. Early guidelines for using multiple views stressed minimizing context switching, as each additional view incurs cognitive cost in preserving the mental map of the data across views \cite{Baldonado2000Guidelines}. Foundational systems like Snap-Together\cite{north2000snap} Visualization introduced user-defined coordination of independent visualizations, enabling synchronized selection and highlighting of data points across views. Such brushing-and-linking techniques, originally conceptualized for 2D plots \cite{becker:1987:BS}, remain a cornerstone: when a user selects an “interesting” subset in one view, corresponding elements in other views are emphasized to maintain contextual awareness of that subset throughout the interface. 



A persistent weakness of early coordinated-view tools was that inter-view relationships were implicit typically a fleeting brush highlight, so users lost track of selections when views were spatially or conceptually distant. Researchers therefore introduced explicit visual links to preserve context. VisLink drew 3-D arcs between related objects in separate 2-D plots, letting analysts trace connections without leaving either view \cite{collins2007vislink}. Cross-application systems extended the idea to multiple windows, overlaying straight link lines on the desktop to guide attention across programs \cite{10.5555/1839214.1839238}. Context-preserving visual links later refined this by routing curved paths around clutter, maintaining each view’s local readability while still tying distant representations together \cite{steinberger2011context}. Together, these advances turned once-fragile highlights into durable, navigable bridges across heterogeneous views.

Beyond simple linking lines, later work softens or removes chart boundaries. Weaver’s cross-filtered views let a drag or slider in one chart instantly filter all others, keeping chosen data visible \cite{weaver2009cross}. Systems such as ConnectedCharts embed small connector graphics \cite{10.1111/j.1467-8659.2012.03121.x}, while MyBrush lets users define their own link sets \cite{8017621}, giving analysts finer control over how views stay synchronized and salient patterns stay in sight.

Recent work focuses on integrating context across views to cut fragmentation. Boundary-blending techniques visibly merge, highlight, or embed elements from one view into another so that noteworthy data remain in sight \cite{sun2023boundaryblendingreconsideringdesign}. Situated analytics applies the same idea in augmented-reality settings, using persistent cues to link physical and virtual views \cite{10536364}. These advances reduce context loss, yet important subsets can still slip away when filters or layouts change gap FluidViews addresses by turning filters into persistent, manipulable tokens.


\subsection{Filtering Techniques in Interactive Visualization}
Interactive filtering enables users to narrow data by hiding clutter or isolating subsets of interest. Early dynamic query systems like FilmFinder and HomeFinder demonstrated the benefits of real-time visual feedback through direct manipulation, tightly coupling input with output to support fluid exploration \cite{10.1145/191666.191775, 545307}. In multi-view environments, filtering often propagates via brushing across coordinated views, a principle generalized by cross-filtering systems to allow iterative, multi-dimensional subset refinement across multiple visualizations \cite{weaver2009cross}.

However, aggressive filtering can obscure users’ memory of prior contexts by removing or muting non-selected items. Focus+context techniques attempt to preserve awareness by softly rendering filtered-out elements instead of erasing them entirely \cite{Heer1998Effects}. Still, basic filtering struggles to maintain visibility of previously identified “interesting” chunks across interaction sequences. Progressive selection models offer a solution: systems like GraphTrail \cite{10.1145/2207676.2208293} record subsets selected through each filtering step as nodes in an analytic trail, enabling backtracking and comparison of intermediate findings. Beyond preserving selections incrementally, contextual bookmarking provides users with explicit means to save and restore analytical states. Contextual snapshots capture selections along with visualization parameters to preserve the original discovery context, allowing restoration even after drastic changes \cite{10.1111/cgf.12406}. Analytic provenance systems such as Voyager \cite{Wongsuphasawat2016VoyagerEA} extend this by automatically logging user interactions, offering undo trails and exploration histories for revisiting earlier analytical paths.

Recent advances have introduced context-aware filtering approaches that dynamically adjust or recommend filters based on prior selections, aiming to preserve notable patterns even as contexts shift. Progressive visual analytics frameworks further reduce the cognitive burden by allowing users to iteratively refine filters while observing partial, real-time results, mitigating the risks of sudden context loss \cite{10373169, Stolper2014Progressive}. Despite these innovations, maintaining awareness of salient data subsets across multi-view exploration and dynamic filtering remains challenging, highlighting the need for mechanisms that persist user-identified “interesting chunks” across shifting analytic contexts.

\section{Designing FluidViews}


MV exploration requires eliminating the high cognitive load and fragmentation common in traditional coordinated-view systems. 
To address the limitations identified in prior systems, we introduce a new interaction paradigm: \textit{Direct Filter Manipulation} (DFM).  
DFM reimagines filters not as static UI controls, but as tangible, manipulable objects embedded within the visualization space. 
This design shift is guided by two key principles, each motivated by cognitive theory and empirical insights from visualization research. 

\textbf{Explicit Persistence (P1)}: Traditional brushing and linking techniques rely on transient interactions hover highlights or ephemeral linkages that disappear after a moment, disrupting multi-step analysis \cite{Baldonado2000Guidelines, becker:1987:BS}. Moreover, filters in many systems are buried in detached control panels or require multiple clicks to reveal or reconfigure. In contrast, we treat filters and highlights as \emph{persistent, visible, and manipulable objects}. This reduces working memory load, externalizes analytic state, and enables users to build upon previous interactions without losing context.

\textbf{Seamless In-Place Filtering (P2)}: Prior work highlights the cognitive cost of interface fragmentation and context switching \cite{pirolli2005sensemaking, hearst2006design}. Systems that require users to navigate modal dialogs, side panels, or separate dashboards for filtering disrupt analytic momentum and inhibit fluid insight generation. Our second principle promotes \emph{in-situ interaction}, where filtering and selection happen directly through gestures and visual marks, embedded in the same space as data exploration. This supports spatial reasoning, analytic flow, and minimizes interface indirection.



To instantiate the DFM paradigm, we introduce four composable, low-friction interaction techniques designed to support core tasks in exploratory analysis filtering, comparison, inspection, and state retention. 
These techniques emphasize direct manipulation, fluid transitions, and embedded interaction. They are:

\textbf{Drag-as-Filter (T1)}: This interaction allows users to pick up any visual mark (e.g., a city on a map or a node in a network) and drop it onto another view to apply a scoped filter(Fig~\ref{fig:teaser}). The dropped object is transformed into a persistent filter token, stored in a visible filter tray, and applied to the target visualization. Multiple marks can be dropped to compose complex, conjunctive filters. This supports rapid, context-preserving filtering across heterogeneous views and reduces dependence on external UI component (P1).

\textbf{Copy-as-Highlight (T2)}: To support lightweight, non-committal comparison, users can copy a visual mark and drop it into another view to trigger a transient highlight(Fig~\ref{fig: copy-as-highlight}). Unlike drag-as-filter, copy-based highlights do not alter the data shown, but instead emphasize relevant subsets without changing the global filter state. These highlights remain until manually removed, supporting revisitation and short-term memory extension (P1, P2).

\textbf{Hover-and-Lock Selection (T3)}: Hovering over a visual mark triggers cross-view highlights in all coordinated views. When a user clicks on the hovered mark, the highlight “locks” in place, making the selection persistent. This mechanism bridges ephemeral exploration with stable state construction, enabling users to incrementally build context while keeping the interface responsive and low-effort (P1).

\textbf{Pop-out (T4)}: For detail-on-demand, any visual element can be right-clicked to spawn a floating micro-view(Fig~\ref{fig:pop-out-microview}). These views present focused information (e.g., detailed timelines, entity relations) without disrupting the main dashboard. Micro-views can be repositioned, layered, and bookmarked, supporting parallel exploration and progressive sensemaking \cite{545307, Perin2014DirectMF} (P1, P2).



\section{FluidViews Technique}


To implement the DFM framework, we developed \emph{FluidViews}, a modular, event-driven multi view system that coordinates interactive operations across heterogeneous visualization types.
This section describes the architectural components, data propagation mechanisms, and drag-and-drop infrastructure that together implement the four primary interaction techniques (T1--T4). 
Our goal is to ensure responsive, low-latency coordination filtering across views, while preserving analytic continuity through shared context and persistent interaction states.

\subsection{Architecture}

FluidViews is built around a modular, extensible workspace that supports four coordinated visualization types representing different views of transactional data: 
(1) a force-directed radial network graph encoding city-to-merchant relationships, 
(2) a geographic heatmap showing state- and city-level aggregates, 
(3) a multipartite Sankey diagram visualizing flows from State \textrightarrow{} City \textrightarrow{} Occupation \textrightarrow{} Merchant, and 
(4) a stacked time-series histogram displaying temporal distributions of transaction volume.

Each view is embedded within a draggable, resizable container, allowing analysts to dynamically arrange their workspace to suit evolving tasks. 
These containers expose standard interaction hooks and subscribe to a global interaction context, enabling views to remain loosely coupled yet tightly synchronized. 
Visualization modules are implemented in D3.js and React, with view updates triggered by changes in shared application state.

\subsection{InteractionContext}

A central architectural element in FluidViews is the \textit{InteractionContext}. A global, observable state container built using React's context system \cite{reactjs}. This shared context serves as the communication bus for all filtering, highlighting, and selection events across the system.

When a user initiates an interaction (e.g., dragging a city node, hovering over a histogram bin), the source visualization generates a structured payload that encodes key metadata about the entity: a unique identifier, type (e.g., city, occupation), origin view, and optionally associated data attributes (e.g., time range, category label). This payload is passed to the \textit{InteractionContext}, which broadcasts it to all subscribed visual components.

Receiving views apply this payload in a view-specific manner. For instance, dragging a city node and dropping it onto the histogram triggers a scoped filter that updates the histogram to show only transactions related to that city. At the same time, the network and Sankey views highlight nodes and paths corresponding to the same city entity. This unified mechanism ensures consistent semantic interpretation of interaction intent, while allowing each view to maintain autonomy over visual encoding.

The event propagation follows a unidirectional flow: user gesture \textrightarrow{} payload generation \textrightarrow{} context update \textrightarrow{} view reaction. This architecture ensures predictable, side-effect-free updates, promotes testability, and supports future extensibility.



\begin{figure}[!htbp]
    \centering
    \includegraphics[width=\linewidth, height=0.6\linewidth]{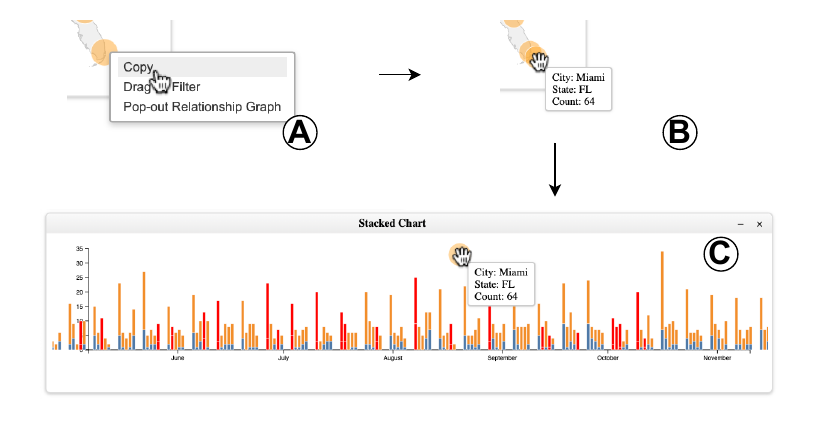}  
    \caption{FluidViews’ \emph{Copy-as-Highlight} workflow.  
(A) The analyst right-clicks a visual mark and chooses \textit{Copy}, creating a draggable duplicate token.  
(B) The token is carried across the visualization while retaining the mark’s data identity.  
(C) Dropping the token onto the stacked time-series chart transiently highlights all bars associated with that entity, enabling rapid cross-view comparison without changing existing filters.}
    \label{fig: copy-as-highlight}
    \vspace{-14pt}  
\end{figure}

\subsection{Drag-and-Drop}

To support (T1) Drag-as-Filter (Fig~\ref{fig:teaser}) and (T2) Copy-as-Highlight (Fig~\ref{fig: copy-as-highlight}), FluidViews implements a unified drag-and-drop infrastructure that binds low-level browser events to high-level analytic actions. Visual objects in all views are equipped with event listeners for \texttt{dragstart}, \texttt{dragover}, and \texttt{drop}, with event handlers registered dynamically based on the visual type and data binding.

\textbf{Drag Initiation}: When a user right-clicks a visual element and selects either \emph{Drag as Filter} or \emph{Copy}, the system creates a \texttt{dragObject} containing the selected mark's metadata: entity ID, entity type, view of origin, and any contextual parameters. This payload is serialized and attached to the browser-native drag event.

\textbf{Drop Handling}: Drop targets in receiving visualizations decode the drag payload and invoke appropriate handlers. For Drag-as-Filter, the payload is interpreted as a filter predicate, which updates the \texttt{InteractionContext}'s selection state. For Copy-as-Highlight, the payload triggers a visual highlight operation without modifying the filter state. In both cases, the receiving view updates immediately, providing real-time feedback and maintaining the illusion of continuity.

\textbf{Filter Persistence}: Drag-as-Filter interactions instantiate persistent tokens in a filter tray, reflecting active filters in the workspace. These tokens remain visible and modifiable, promoting user awareness and traceability of active constraints. While not a standalone interaction, this design complements the drag interaction by preserving its analytic footprint.

\subsection{Hover-Highlight and Click-to-Lock}

To support (T3) Hover-and-Lock Selection, all visual marks in FluidViews expose hover event bindings. Hovering over a visual element (e.g., a Sankey node or heatmap cell) dispatches a highlight payload to the \textit{InteractionContext}, which then propagates to other subscribed views. The system uses soft highlighting (e.g., opacity changes, stroke overlays) to draw attention to semantically related elements without disrupting layout.

To transition from transient exploration to persistent comparison, users can click on a hovered item to "lock" the highlight. This promotes analytic continuity by allowing users to accumulate focus without needing to re-hover. Locked selections persist until explicitly cleared, supporting multi-step comparisons and enabling advanced inspection workflows.

The hover-lock interaction model is lightweight and non-intrusive, providing a middle ground between traditional brushing and more heavyweight filter construction. It encourages rapid exploration while giving users fine-grained control over attention management.



\begin{figure}[t]
  \centering
  \includegraphics[width=\linewidth, height=0.6\linewidth]{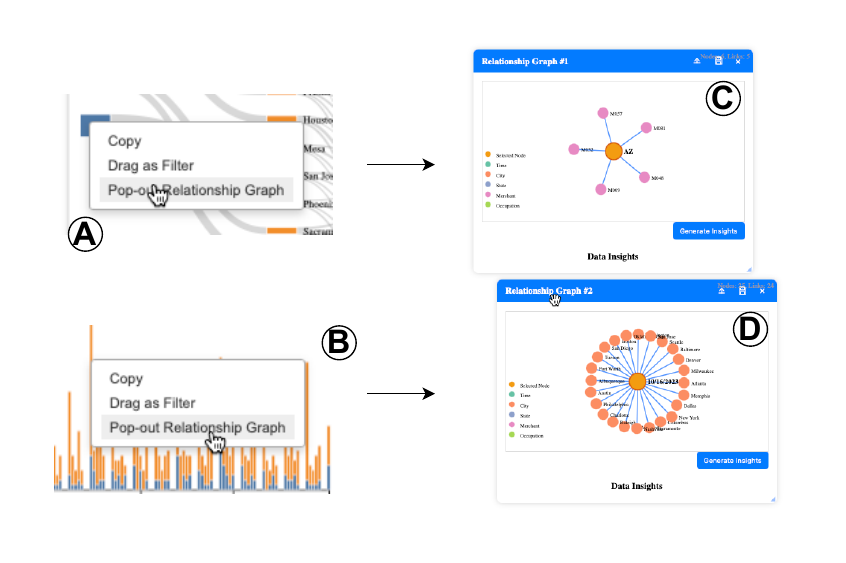} 
  \caption{
Pop-out micro-views in FluidViews.  
(A) From the Sankey diagram, the user invokes the Pop-out micro-views to inspect a selected occupation, triggering a focused micro-view (C) that visualizes first-hop entity relationships.  
(B) Similarly, from the stacked time-series chart, a right-click on a visual mark spawns a pop-out graph (D) showing temporal connections for the selected city.  
These floating, spatially independent micro-views allow analysts to perform targeted inspections without disrupting the main dashboard context.
}
  \label{fig:pop-out-microview}
  \vspace{-13pt}
\end{figure}

\subsection{Pop-out Micro-Views for Detail-on-Demand}

FluidViews also supports (T4) Pop-out Micro-Views to enable focused, context-preserving inspection . Right-clicking on any visual mark opens a context menu with the option to spawn a micro-view (Fig~\ref{fig:pop-out-microview}). These pop-outs are floating containers that render secondary views such as zoomed-in timelines, entity-specific Sankey flows, or localized heatmaps based on the selected data entity. Micro-views are spatially independent and persist across dashboard interactions, enabling analysts to construct side-by-side comparisons, stage insights, or bookmark specific threads of analysis. Each micro-view is linked to its originating entity and updates dynamically if the underlying data changes. 

Users can reposition, resize, and layer these containers to suit complex exploratory workflows. This design extends classic detail-on-demand interactions \cite{545307}, while supporting modern analytic behaviors such as parallel investigation, multi-entity tracking, and iterative drill-down.

\section{Evaluation}
To assess the effectiveness and usability of FluidViews, we conducted a qualitative user study focusing on how the system supports multi-view exploration and coordination tasks. Three participants with prior experience in data analysis and visual interfaces participated: a professional data scientist, an HCI researcher, and a graduate student specializing in visualization. Each session lasted approximately 45 minutes and included a structured introduction, hands-on exploration of a multi-dimensional transactional dataset, and a post-session semi-structured interview.

Participants were presented with three core tasks reflecting realistic exploratory goals: (1) identify regional variations in merchant activity using the geographic heatmap and time-series view, (2) trace occupation-based flows in the Sankey diagram and correlate with merchant types, and (3) investigate time-based anomalies for specific cities using pop-out micro-views. All participants were encouraged to verbalize their thought process, and their interactions and comments were recorded for subsequent analysis.

Participants quickly adapted to the interface, particularly the drag-and-drop filtering and hover-based selection mechanisms. 
No formal training was needed beyond a two-minute walkthrough. 
All users noted that persistent filter tokens and copy-as-highlight helped them layer and compare multiple analytic hypotheses. These artifacts served as visual anchors and supported tracking across views. One participant said, "Even after five steps, I didn’t lose track of what I filtered, everything was there in front of me."
Participants frequently used a combination of hover-to-lock and drag-as-filter to probe hypotheses, lock insights, and propagate context. This coordination enabled fast, iterative discovery. 
For example, one participant locked highlights of two occupations, then filtered by city to compare distributional effects across time. 
The micro-view functionality was used extensively to isolate and inspect sub-patterns. 
Two participants created multiple micro-views in parallel, aligning them to compare temporal shifts between entities. 
This emergent behavior suggests the value of the design in supporting nonlinear exploration paths. 
Despite overall positive reception, participants identified several challenges. 
Two noted the absence of an \emph{undo} action after misplaced filters. 
Managing multiple micro-views became cumbersome without layout snapping or grouping support. One participant suggested a command center or summary view to monitor all active filters and highlights.

\section{Discussion and Limitations}

The results of the preliminary study support our core design rationale: that embedding filtering and selection within the views, and making these interactions persistent and manipulable, can improve users’ ability to construct, coordinate, and refine complex queries across multiple dimensions. The participants’ ability to discover and adopt new interaction sequences without training indicates that the DFM approach has strong potential for supporting sensemaking in visual analytics environments.

The study, however, is subject to limitations. With only three participants and a single dataset, the findings are exploratory and not statistically generalizable. All users had at least moderate prior experience with interactive visualizations, which may have positively biased usability perceptions. The evaluation did not measure task performance quantitatively; future work should include comparative timing, accuracy, and cognitive load assessments.

Moving forward, we plan to conduct larger controlled studies comparing FluidViews to traditional coordinated systems that rely on static side panels and filter dialogs. We are also exploring mechanisms for adaptive view layouts, filter grouping, and collaborative provenance trails to enhance traceability and group analysis scenarios. Ultimately, we believe that the Direct Filter Manipulation framework opens a promising direction for building more expressive and context-aware analytic interfaces.

\section{Conclusion}
FluidViews introduces a direct-manipulation framework for multiview visual analytics, enabling users to construct persistent filters, trigger transient highlights, and access on-demand micro-views through intuitive drag-and-drop gestures. By embedding these interactions directly within the visualization space, our approach addresses key limitations of conventional systems by reducing reliance on external filter panels, minimizing context switches, and supporting more fluid cross-view exploration. Future work will extend FluidViews to new analytic domains and explore adaptive layouts for multi-token interactions.

\bibliographystyle{abbrv-doi}

\bibliography{template}
\end{document}